# Synthesis and Superconductivity in yttrium superhydrides under high pressure[*]

Yingying Wang[1,2], Kui Wang[1,2], Yao Sun[1,2], Liang Ma,[1,2,3] Yanchao Wang,[1,2] Bo Zou,[1] Guangtao Liu[2†], Mi Zhou[2‡] and Hongbo Wang[1,2§]

[1] *State Key Laboratory of Superhard Materials, College of Physics, Jilin University, Changchun 130012, China*
[2] *International Center of Computational Method & Software, College of Physics, Jilin University, Changchun 130012, China*
[3] *International Center of Future Science, Jilin University, Changchun 130012, China*

[†] Corresponding author. E-mail: liuguangtao@jlu.edu.cn
[‡] Corresponding author. E-mail: mzhou@jlu.edu.cn
[§] Corresponding author. E-mail: whb2477@jlu.edu.cn

The flourishing rare earth superhydrides are a class of recently discovered materials that possess near-room-temperature superconductivity at high pressures, opening a new era of superconductivity research at high pressures. Among these superhydrides, yttrium superhydrides attracted great interest owing to their abundance of stoichiometries and excellent superconductivities. Here, we carried out a comprehensive study of yttrium superhydrides in a wide pressure range of 145-300 GPa. We successfully synthesized a series of superhydrides with the compositions of $YH_4$, $YH_6$, $YH_7$, and $YH_9$, and reported their superconducting transition temperatures of 82 K at 167 GPa, 218 K at 165 GPa, 29 K at 162 GPa, and 230 K at 300 GPa, respectively, which were evidenced by a sharp drop of resistivity. The structure and superconductivity of $YH_4$, which was taken as a representative example, were also examined by X-ray diffraction measurements and the suppression of the superconductivity under external magnetic fields, respectively. Clathrate $YH_{10}$ as a candidate of room-temperature superconductor was not synthesized within the studied pressure and temperature ranges of up to 300 GPa and 2000 K, respectively. The current work created a detailed platform for further searching room-temperature superconductors in polynary yttrium-based superhydrides.

## 1. Introduction

The search for high-temperature superconductors (HTS) with superconducting transition temperature ($T_c$) above liquid-nitrogen temperature has long been recognized as an interesting topic since the discovery of Hg with $T_c$ = 4.2 K.[1] According to the

[*] Project supported by the National Key Research and Development Program of China (Grant No. 2021YFA1400203 and No. 2018YFA0305900), National Natural Science Foundation of China (Grant No. 52090024, No. 11874175, No. 12074139, No. 12074138, No. 11874176, and No. 12034009), the Strategic Priority Research Program of Chinese Academy of Sciences (Grant No. XDB33000000), and Program for JLU Science and Technology Innovative Research Team (JLUSTIRT).

Bardeen-Cooper-Schrieffer theory,[2] metallic hydrogen (MH) is one of the best candidates for achieving HTS; however, the quest for MH has been proven extremely challenging due to the requirements of ultrahigh pressures conditions. In 1970, Satterthwaite et al. found ~8 K superconductivity in thorium hydride, suggesting that hydrogen-rich metal hydrides would expect to be HTS.[3] Then, Gilman[4] and Ashcroft[5] further proposed that MH could be achieved in hydrogen-rich hydrides at lower pressures since the heavier atoms played a chemical precompression on hydrogen, opening up an era of the search for HTS in hydrogen-rich compounds at high pressures. However, despite considerable efforts, there has been no experimental breakthrough for a long time until the observation of 203 K superconductivity at 155 GPa in covalent $H_3S$,[6] which further inspired the search for HTS in conventional phonon-mediated hydrides superconductors.

In contrast to covalent superhydrides such as $H_3S$, ionic metal hydrides provide more choices in searching for HTS. In 2012, Wang et al. predicted the first clathrate hydride of $CaH_6$ with a very high $T_c$ of 235 K at 150 GPa.[7] Following this work, a long list of clathrate $REH_6$, $REH_9$, and $REH_{10}$ hydrides (RE: rare earth metal) were also predicted to possess high $T_c$ values approaching or even above room temperature.[8-10] Stimulated by these predictions, a series of clathrate superhydrides, such as $CaH_6$,[11] $LaH_{10}$,[12,13] $CeH_9$, $CeH_{10}$,[14] $ThH_9$, $ThH_{10}$,[15] $(La, Y)H_{10}$,[16] etc., were successfully synthesized with observed $T_c$ ranging from 57-260 K. Among ionic superhydrides, yttrium superhydrides attracted intense interest due to their abundant stoichiometries, which were predicted to possess high superconducting transition temperatures, e.g., 84-95 K at 120 GPa in $YH_4$,[10,17] 251-264 K at 120 GPa in $YH_6$,[10] 32-43 K at 165 GPa in $YH_7$,[18] 253-276 K at 150 GPa in $YH_9$[8] and 305-326 K at 250 GPa in $YH_{10}$ [9]. Recently, Kong et al. [19] successfully synthesized $YH_4$ and clathrate structured $YH_6$ and $YH_9$ with the observed $T_c$s of 220 K at 183 GPa and 243 K at 201 GPa for the last two yttrium superhydrides, respectively. Meanwhile, Troyan et al. also independently synthesized a clathrate $YH_6$[18], with an observed $T_c$ of 224 K at 166 GPa. After that, Snider et al. synthesized $YH_9$ with a $T_c$ up to 262 K by catalytic hydrogenation at 182 GPa.[20] Furthermore, recent research successfully observed an 88 K superconductivity of $YH_4$ at 155 GPa.[21]

Besides binary yttrium superhydrides, yttrium-bearing ternary hydrides, where the introduction of a third element other than hydrogen greatly expands the phase space, have attracted extensive attentions. Liang et al.[22] and Xie et al.[23] predicted a clathrate $CaYH_{12}$ with an estimated $T_c$ of 258 K at 200 GPa and 230 K at 180 GPa, respectively. Then, Liang et al. predicted a ternary $YSH_6$ with a $T_c$ of 91 K at 210 GPa.[24] Experimentally, $(La, Y)H_6$ and $(La, Y)H_{10}$[16] were synthesized at high pressures with $T_c$s of 237 K and 253 K, respectively.

Previously, each work only focused on the superconductivity of one or two hydrides, even though more superhydrides were synthesized. So far, there is still a lack of efforts to systematically investigate the superconductivity of all the experimentally reported unconventional superhydrides. In this work, we first carried out the detailed structural and superconductivity studies of $YH_4$, which was taken as an example due to its rare investigation previously. X-ray diffraction measurements revealed the successful synthesis of predicted *I4/mmm*-YH$_4$ at about 167 GPa and 1600 K, and its measured $T_c$ of 82 K was evidenced by a sharp drop of the resistivity and a characteristic decrease of superconducting transition under a magnetic field up to 8.5 T. The further electrical transport measurements revealed a series of additional superconductivities of 29 K (162 GPa), 218 K (165 GPa), and 230 K (300 GPa), which arise from YH$_7$ and clathrate structured YH$_6$, and YH$_9$, respectively, inferred from the consistency of $T_c$s with previous studies.

2. **Experiment methods**

According to the different target pressures, symmetric diamond anvil cells (DACs) outfitted diamond anvils with a culet size of ~30-60 μm beveled at 8.5° to a diameter of ~250 μm. The composite gasket was composed of rhenium outer annulus and a mixture of epoxy resin and Al$_2$O$_3$ powder. The insulating gasket was pre-indented to 10 μm in thickness, and the corresponding sample chamber with a diameter of 20-30 μm was drilled by a laser drilling system. Commercially available yttrium ingot (Alfa Aesar, 99.9% purity) and NH$_3$BH$_3$ (AB) powder (Sigma-Aldrich, 97%) were loaded into the sample chamber inside a glovebox filled with Ar atmosphere with O$_2$ and H$_2$O contents of < 0.01 ppm. The Y foil and Au electrodes with the thicknesses of 2 μm and 1 μm, respectively, were sandwiched between the AB. Subsequently, the samples were compressed to the required synthesis pressure before laser heating. The pressure was calibrated by the high-frequency edge of the diamond Raman line[25]. The laser heating of the sample was performed using a pulsed YAG infrared laser, and the temperature was determined using the blackbody radiation fit within the Planck function. *In-situ* high-pressure angle-dispersive x-ray diffraction (ADXRD) experiments were performed at BL15U1 beamline of the Shanghai Synchrotron Radiation Facility (5×12 μm$^2$) using a monochromatic beam wavelength of 0.6199 Å and the average acquisition time was 120 s. Before the experiment, the relevant geometric parameters were calibrated using a CeO$_2$ standard. A Mar165 CCD detector was used to collect diffraction patterns and the collected patterns were analyzed using DIOPTAS software, yielding one-dimension profiles.[26] The Le Bail profile matching refinements were performed using the GSAS + EXPGUI programs.[27] Based on the four-probe van der Pauw method,[28] the resistance measurements were performed with currents of $10^{-6}$–$10^{-4}$ A (Keithley 2182A nanovoltmeter and 6221 AC and DC current source) and the selected data were the warming cycles with a controlled rate of about 1 K min$^{-1}$. In addition, non-magnetic DACs made of Be-Cu alloy were used for resistance measurements under an external magnetic field up to 8.5 T.

3. **Results and discussion**

In this work, we prepared 11 samples, designated as sample_1 through sample_11, to synthesize yttrium superhydrides from a mixture of Y and AB, and to explore their superconductivities. AB has been demonstrated to be a reliable H$_2$ source from previous excellent results.[11,13,18,19,21] At high temperatures, AB would decompose into H$_2$ plus *c*-BN, the latter avoiding the problem of poor contact between the synthesized product and electrodes. The schematic diagram of the assembly used for synthesis and subsequent four-probe electrical resistance measurements is shown in Fig. 1(a). In sample_1, the reactants were initially compressed to 167 GPa (Fig. S1(a)) and then heated to about 1600 K. The clear H-H vibration (Fig. S1(b)) demonstrates a hydrogen-rich environment. The sample turned black after laser heating indicating a chemical reaction occurred (inset in Fig. 1(b)). Representative electrical resistance measurements as a function of temperature at high pressures clearly showed a superconducting transition as evidenced by the sharp drop of the resistance at 82 K, as displayed in Fig. 1(b). This superconducting transition can be perfectly reproduced in several independent experiments (Fig. S2), further confirming the reliability of our results. To determine the highest value of $T_c$, we evaluated the pressure dependence of $T_c$, as shown in Fig. 3(b). $T_c$ fluctuates slightly in the pressure range of 145-170 GPa in different experiments and the highest $T_c$ of 86 K at 160 GPa is consistent with the previous theoretical estimation of 74-95 K at the same pressure for $YH_4$. With decreasing pressure, the superconducting transition disappeared at about 143 GPa (Fig. S2), indicating a possible decomposition of superconducting phase.

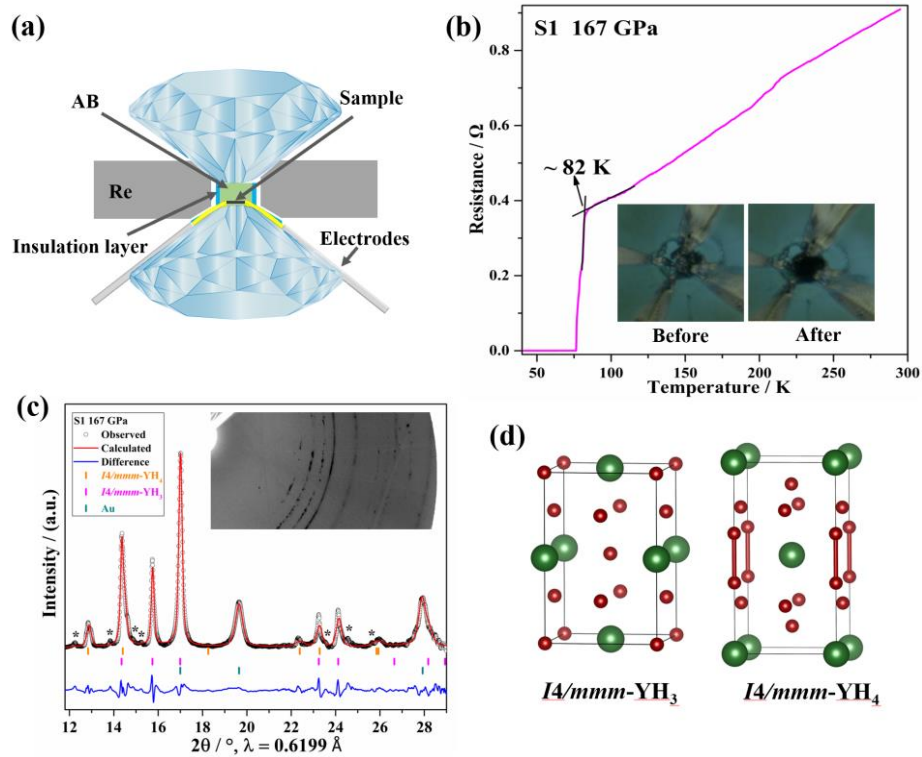

FIG. 1. (a) Schematic of the experimental setup for synthesis and four-probe superconducting electrical resistance measurements. (b) Temperature dependence of resistance in sample_1 (S1) at 167 GPa. The insets show optical micrograph of the sample before and after laser heating. The value of the $T_c$ is defined as the crossing point of the resistance slopes before and after the resistance drop. (c) Synchrotron XRD pattern of yttrium hydrides in S1 at 167 GPa. The inset displays the two-dimensional XRD pattern. Unidentified weak reflections are marked by asterisks. (d) Crystal structures of *I4/mmm*-$YH_3$ and *I4/mmm*-$YH_4$. Big and small balls represent Y and H atoms, respectively.

To further determine the structure of the high-temperature superconducting phase, we performed *in-situ* high-pressure ADXRD measurements of sample_1, which revealed that the products were dominated by *I4/mmm*-YH$_3$ and *I4/mmm*-YH$_4$ as shown in Fig. 1(c) and the refined structural information were listed in Table S1. The tetragonal YH$_3$, possessed a new high-pressure phase besides the conventional fcc phase, was synthesized for the first time after prediction.[17] And no superconductivity was predicted in *I4/mmm*-YH$_3$ up to 200 GPa. Therefore, the observed superconducting transition in sample_1 should originate from YH$_4$.

To date, the measurement of the Meissner effect in ultra-high-pressure experiments remains a great challenge due to the small sample size. An applied external magnetic field could break the Cooper pairs, thus reducing the value of $T_c$; therefore, the suppression of superconducting transitions by an applied magnetic field can be used to analyze the nature of the superconducting states. Fig. 2(a) shows the measured resistance of sample_2 under different magnetic fields at 170 GPa. It is obvious that the $T_c$ decreased from 77 K to 53 K as the magnetic field increased to 8.5 T indicating the superconducting nature of the transition. As shown in Fig. 2(b), the extrapolated upper critical field $\mu_0H_{c2}(T)$ and coherence length toward T = 0 K are 14.9 T and 47 Å and 18.7 T and 42 Å were fitted by the Ginzburg–Landau (GL)[29] and Werthamer-Helfand–Hohenberg (WHH)[30] models, respectively. Furthermore, in this experiment, besides the superconductivity of YH$_4$, we also observed another low-temperature superconductivity of 17 K (inset in Fig. 2(a)), which can be attributed to element yttrium according to the consistency of $T_c$ in the unheated sample (Fig. S4). Similar results for YH$_4$ were independently reported by another group.[21]

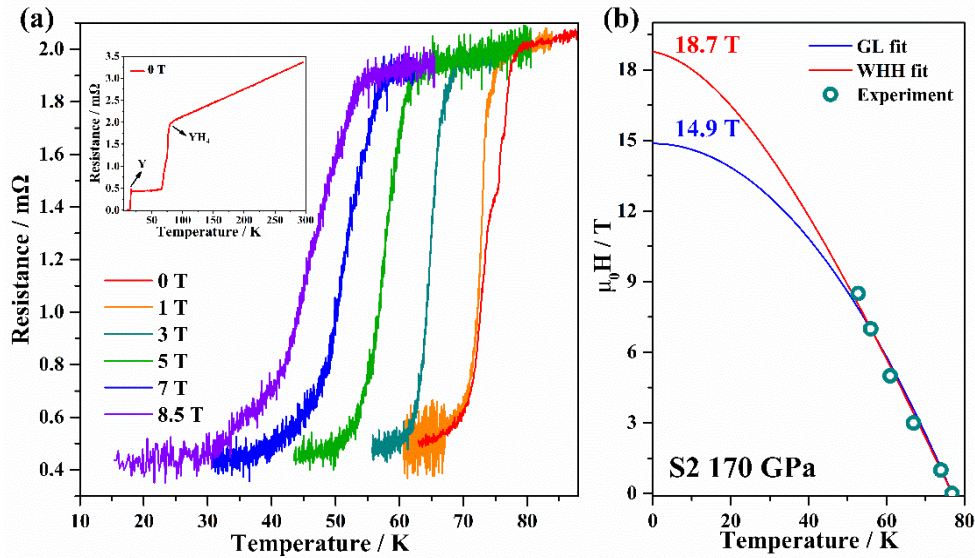

FIG. 2. (a) The temperature dependence of the resistance for the *I4/mmm*-YH$_4$ under external magnetic fields of H = 0, 1, 3, 5, 7 and 8.5 T at 170 GPa in sample_2 (S2). Inset: the temperature-resistance curve without external magnetic fields. (b) The fitted upper critical magnetic field by GL and WHH models.

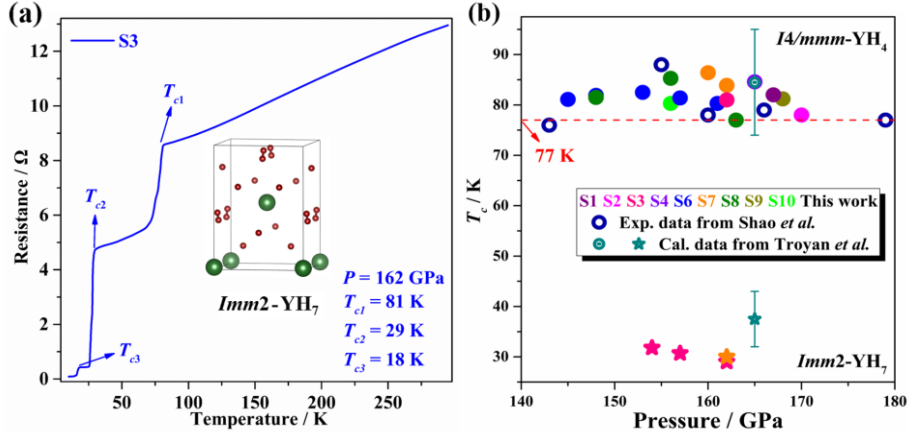

FIG. 3. (a) Temperature dependence of resistance in sample_3 (S3) at 162 GPa. Inset: crystal structures of $Imm2$-$YH_7$. Big and small balls represent Y and H atoms, respectively. (b) The pressure dependence of $T_c$ for $I4/mmm$ $YH_4$ (circle) and $Imm2$-$YH_7$ (star). Different colors represent different samples. The cited experimental data for $YH_4$ are represented by open circles.[21] Dark cyan symbols depict the calculated data from Troyan et al.[18]

Furthermore, after laser heated the sample_3 to about 1750 K at 162 GPa, we observed step-down behaviors at 81 K, 29 K, and 18 K in electrical resistance measurement (Fig. 3(a)). As discussed above, the first and third resistance drops arise from the superconducting transitions of $YH_4$ and element Y, respectively. Besides, a new superconducting transition was observed at 29 K, which has never been reported. Based on previous theoretical work,[18] we speculated that the second resistance drop may originate from $Imm2$-$YH_7$ and this superconductivity transition was also observed in sample_7 (Fig. S2). The pressure dependencies of $T_c$ for $YH_7$ and $YH_4$ were summarized in Fig. 3(b). Similar to the variation trend of $YH_4$, the $T_c$ of $YH_7$ was relatively stable in the pressure range of 154-170 GPa. Although both $I4/mmm$-$YH_4$ and $Imm2$-$YH_7$ have molecular "$H_2$" unit (Fig. 1(d) and Fig. 3(a)), the $T_c$ of $YH_4$ with high-symmetry structure is higher than that of $YH_7$ due to the stronger electron-phonon coupling.[18]

In the following work, we tuned the heating temperature and pressure, aiming to synthesize the high-temperature superconducting clathrate $YH_6$, $YH_9$, or even $YH_{10}$. For sample_4, when we elevated the heating temperature to 2200 K at 165 GPa, a superconducting temperature of 218 K was observed (Fig. 4(a)). Subsequently, the sample_5 was compressed to a superhigh pressure of 300 GPa (Fig. S1(a)) and heated to about 2000 K, and the electrical resistance measurement curve revealed a superconductivity at 230 K (Fig. 4(a)). It can be seen from Fig. 4(b) that the $T_c$s of samples_4 and samples_5 perfectly match the reported experimental results for clathrate structured $YH_6$ and $YH_9$,[18,19] respectively. The high $T_c$ of $Im\bar{3}m$-$YH_6$ and $P6_3/mmc$-$YH_9$ were attributed to their hydrogen cage structure, and especially to the large contribution of the H derived electronic density of states at the Fermi level.[8,10] Unfortunately, we did not observe any sign of clathrate $YH_{10}$, which may be synthesized at higher pressures.

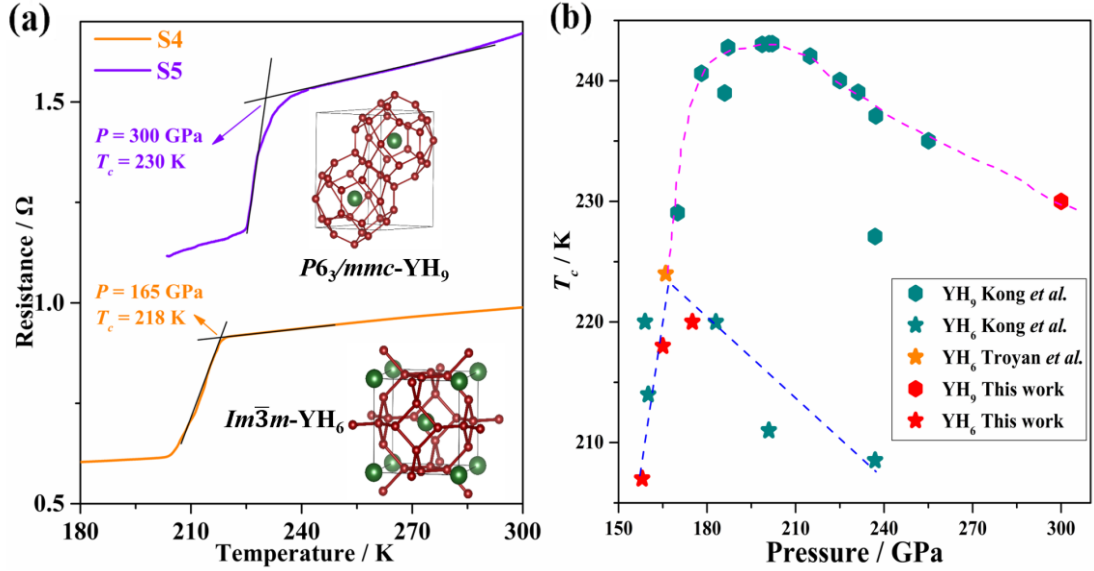

FIG. 4 (a) Temperature dependence of resistance in sample_4 (S4) at 165 GPa and sample_5 (S5) at 300 GPa. Inset: crystal structures of $Im\bar{3}m$-$YH_6$ and $P6_3/mmc$-$YH_9$. Big and small balls represent Y and H atoms, respectively. (b) The pressure dependence of $T_c$ for $Im\bar{3}m$-$YH_6$ (star) and $P6_3/mmc$-$YH_9$ (hexagon). The symbols of dark cyan, orange and red correspond to the data of Kong et al.,[19] Troyan et al.[18] and this work, respectively.

A series of superhydrides with high $T_c$ have been synthesized under high pressures; however, the absence of the broadening of resistive transition with increasing magnetic field in some work[12,16,31] led to an argument about their superconductivities.[32] Taking $YH_4$ as an example, we observed a clear broadening of the resistive transition under applied magnetic fields (Fig. S3), which show a similar trend to those of typical standard superconductors, such as $MgB_2$[33] and NbN,[34] further demonstrating the reliability of our results. As a member of the superhydrides, the results of resistive transport measurement under external magnetic fields in $YH_4$ will contribute to the understanding of the debate on the superconductivity in superhydrides.

## 4. Conclusion and perspectives

In summary, we have successfully synthesized $YH_4$, $YH_6$, $YH_7$, and $YH_9$, that exhibits $T_c$s of 82 K at 167 GPa, 218 K at 165 GPa, 29 K at 162 GPa, and 230 K at 300 GPa, respectively. Furthermore, a tetragonal phase as a new high-pressure structure of conventional $YH_3$ was synthesized for the first time at 167 GPa. These results confirmed the original theoretical prediction and provided a platform for further exploration of high-temperature superconductors on the doped Y-based polynary superhydrides.

This paper is dedicated to the 70th anniversary of the physics of Jilin University.

**Acknowledgment**

XRD measurements were performed at BL15U1 station in Shanghai Synchrotron Radiation Facility (SSRF) and 4W2 station in Beijing Synchrotron Radiation Facility (BSRF). The measurements of superconducting transition under external magnetic fields were supported by the Synergic Extreme Condition User Facility (SECUF) and China's Steady High Magnetic Field Facility (SHMFF).

# Supplemental Materials

# Synthesis and Superconductivity in yttrium superhydrides under high pressure

Yingying Wang[1,2], Kui Wang[1,2], Yao Sun[1,2], Liang Ma[1,2,3], Yanchao Wang[1,2], Bo Zou[1], Guangtao Liu[2†], Mi Zhou[2‡] and Hongbo Wang[1,2§]

[1] *State Key Laboratory of Superhard Materials, College of Physics, Jilin University, Changchun 130012, China*
[2] *International Center of Computational Method & Software, College of Physics, Jilin University, Changchun 130012, China*
[3] *International Center of Future Science, Jilin University, Changchun 130012, China*

† Corresponding author. E-mail: liuguangtao@jlu.edu.cn
‡ Corresponding author. E-mail: mzhou@jlu.edu.cn
§ Corresponding author. E-mail: whb2477@jlu.edu.cn

Table S1 The lattice parameters and volumes of $I4/mmm$-YH$_3$ and $I4/mmm$-YH$_4$.

| | | $I4/mmm$-YH$_3$ | | |
|---|---|---|---|---|
| DAC | Pressure (GPa) | a (Å) | c (Å) | V (Å$^3$) |
| S1 | 158 | 2.945 | 4.497 | 39.00 |
| Cal.[a] | 150 | 2.995 | 4.329 | 38.83 |
| | | $I4/mmm$-YH$_4$ | | |
| DAC | Pressure (GPa) | a (Å) | c (Å) | V (Å$^3$) |
| S1 | 158 | 2.769 | 5.498 | 42.155 |
| Exp.[b] | 147 | 2.769 | 4.961 | 38.03 |
| Exp.[a] | 168 | 2.751 | 5.150 | 39.01 |
| Cal.[a] | 150 | 2.799 | 5.278 | 41.35 |

[a] Ref. [1]

[b] Ref. [2]

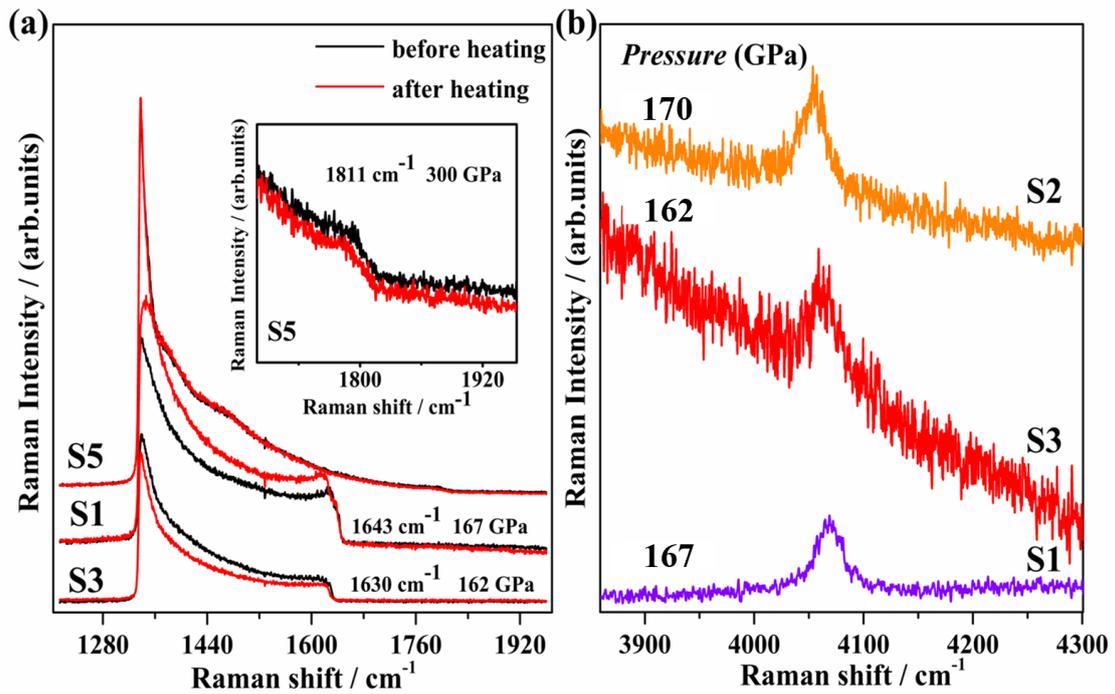

FIG. S1 Typical Raman spectra of diamonds (sample_1, sample_3 and sample_5) and H$_2$ (sample_1, sample_2 and sample_3).

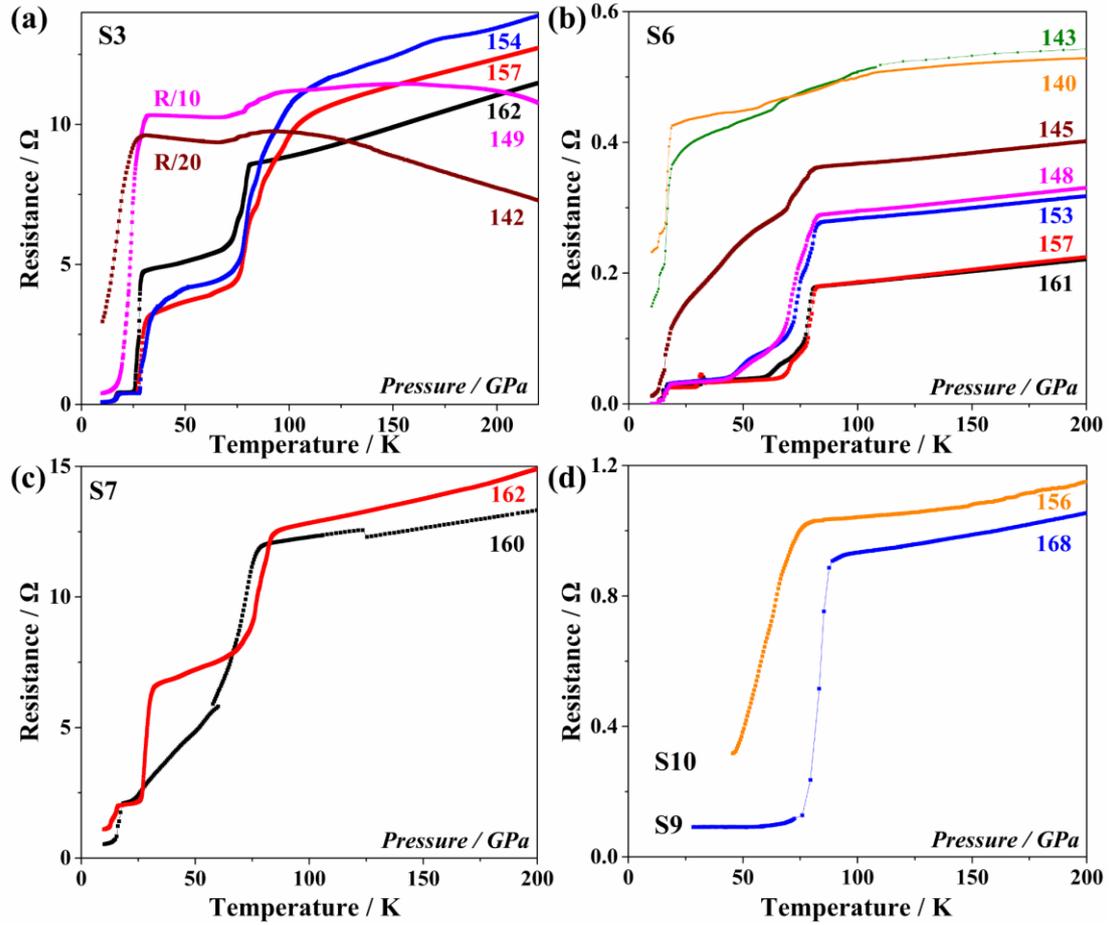

FIG. S2 Temperature dependence of resistance of sample_3 (a), sample_6 (b) and sample_7 (c) measured upon decompressions. Three samples were synthesized at around 160 GPa from a mixture of Y and AB. (d) Temperature dependence of resistance of Y-H samples in sample_9 and sample_10 synthesized at 168 and 156 GPa, respectively.

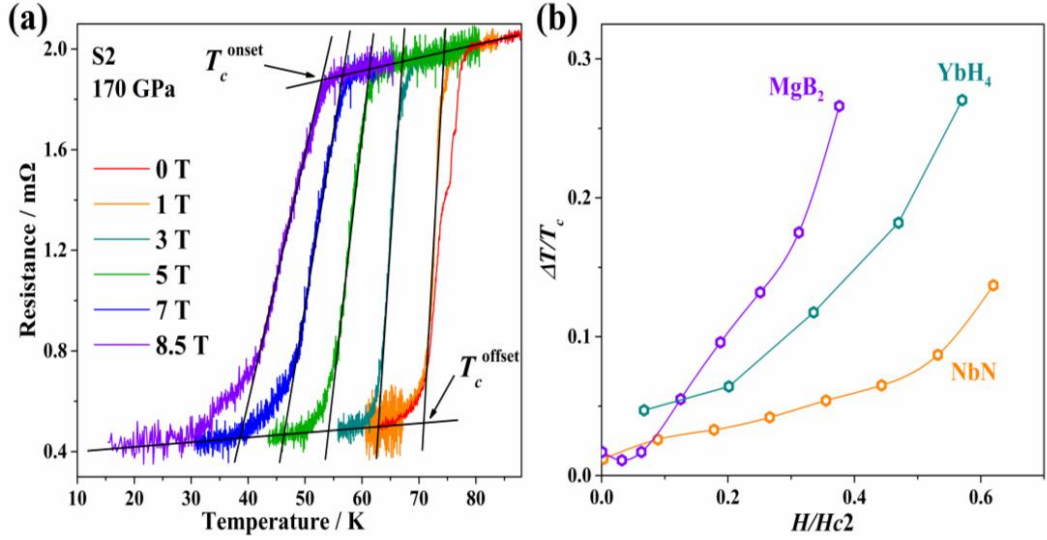

FIG. S3 (a) Temperature dependence of the resistance for the *I4/mmm*-YH$_4$ under external magnetic fields at 170 GPa in sample_2. The broadening of the transition is defined as $\Delta T = T_c^{onset} - T_c^{offset}$, where $T_c^{onset}$ and $T_c^{offset}$ are defined as the intersection of the extension line of the transition curve and the extension line before and after the superconducting transition. (b) Comparison of the field dependence of the superconducting transition widths for YH$_4$, MgB$_2$[3] and NbN[4]. Obviously, a clear broadening of the resistive transition of YH$_4$ under applied magnetic fields was observed, which is similar to the trend of standard superconductors. It should be noted that since the resistance of YH$_4$ does not drop sufficiently sharply at 0 T, the superconducting transition width versus field is plotted starting from 1 T in order to minimize the error.

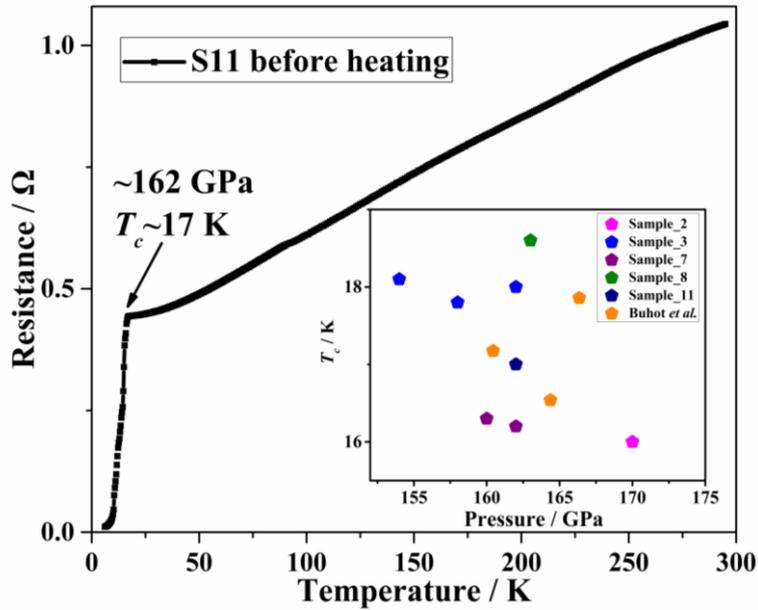

FIG. S4 Temperature dependence of the resistance of sample_11 at 162 GPa before laser heating. Superconductivity with a $T_c$ ~17 K is consistent with that of element yttrium.[5] The pressure dependence of the superconductivity is shown in the inset. Orange symbols depict the experimental data from Buhot et al.[6]